# Human-Centric Decision Support Tools: Insights from Real-World Design and Implementation

Narges Ahani[1] and Andrew C. Trapp[1,2]

[1]Data Science Program, Worcester Polytechnic Institute, Worcester, MA

[2]Foisie Business School, Worcester Polytechnic Institute, Worcester, MA

**Abstract**: Decision support tools enable improved decision making for challenging decision problems by empowering stakeholders to process, analyze, visualize, and otherwise make sense of a variety of key factors. Their intentional design is a critical component of the value they create. All decision-support tools share in common that there is a complex decision problem to be solved for which decision-support is useful, and moreover that appropriate analytics expertise is available to produce solutions to the problem setting at hand. When well-designed, decision support tools reduce friction and increase efficiency in providing support for the decision-making process, thereby improving the ability of decision-makers to make quality decisions. On the other hand, the presence of overwhelming, superfluous, insufficient, or ill-fitting information and software features can have an adverse effect on the decision-making process and, consequently, outcomes. We advocate for an innovative, and perhaps overlooked, approach to designing effective decision support tools: genuinely listening to the project stakeholders, to ascertain and appreciate their real needs and perspectives. By prioritizing stakeholder needs, a foundation of mutual trust and understanding is established with the design team. We maintain this trust is critical to eventual tool acceptance and adoption, and its absence jeopardizes the future use of the tool, which would leave its analytical insights for naught. We discuss examples across multiple contexts to underscore our collective experience, highlight lessons learned, and present recommended practices to improve the design and eventual adoption of decision support tools.

## I. The Increasing Prevalence and Importance of Decision Support Tools

Rapid advances in information technology are enabling the collection of increasingly vast amounts of data more quickly and easily than ever before (Hoch and Schkade 1996). At the heart of decision support tools lies *analytics*, the systematic computational analysis of data and statistics to inform decision making, which can be descriptive, predictive, or prescriptive in nature (Davenport and Harris 2017). While the presence of more and better information can empower better decisions, our elevated access to nearly inconsumable amounts of data does not alone guarantee better decision-making. The human mind is limited in available processing power; a key study found that decision makers are unable to identify nearly half of the attractive options (Siebert and Keeney 2015). Information at hand is often only marginally relevant, and the presence of overwhelming, superfluous, and partial information only complicates the decision-making process.

There is an increasing need for tools and systems that effectively analyze available data and inform decision makers with fact-based, data-driven insights. Decision makers not only want to find the best solution – they also want it *quickly*. Decision support tools are computer-based technologies that facilitate better decision-making by solving complex problems and enabling human interaction (Shim et al. 2002). The main aim of decision support tools is to provide decision makers with technology that enhances their capability of decision-making, resulting in making more informed decisions (Arnott and Pervan 2008). Well-designed decision support tools improve the quality of decisions on important issues by removing friction and increasing efficiency in problem-solving. Such systems alleviate the condition of information overload by presenting the right information at the right time, thereby boosting decision-making effectiveness.

Various domain knowledge and associated technologies have been incorporated in decision support tools including Artificial Intelligence, Business Intelligence, Decision Sciences, Machine Learning, Operation Research, Psychology, User Experience, and related fields. Many decision support systems combine knowledge and technologies from multiple domains to form an integrated tool to aid in resolving decision problems specific to a certain set of stakeholders. As technology continues to evolve, data-driven decision support systems have advanced in sophistication and application to new and exciting areas. Throughout this study, we refer to *designers* as the role primarily involved with the creation of the decision support technology, and *stakeholders* as the general role representing clients, end-users, decision-makers, and their management – really, anyone who is involved in the decision-making process, recognizing that these roles vary from organization to organization.

## II. Characteristics of an Effective Decision Support System

While decision support tools hold great promise, not all decision support tools have a successful story to tell. Many projects were launched to design and develop a decision support tool for a specific decision-making context, but ultimately failed because the final product was not *successfully adopted* by key stakeholders (Pynoo et al. 2013, Bhattacherjee and Hikmet 2007, Freudenheim 2004, Briggs and Arnott 2001, Rainer and Watson 1995, Hurst et al. 1983). In this chapter we focus on factors that lead to successful tool adoption, the most important of which is to design with the purpose of aligning with the needs of key stakeholders. There exists a tendency – perhaps understandably so – for designers to overly focus on the development of decision-making models and algorithms; in so doing, this may compromise the ability to recognize and satisfy the exact needs of decision makers. While cutting-edge algorithms certainly have their place, only by sufficiently aligning with stakeholder needs does any project have the opportunity to succeed. Regardless of the level of technical sophistication, in the end the effectiveness and value of the tool largely depends on the extent to which it will be adopted and put into practice by practitioners and decision-makers (Gönül, Önkal, and Lawrence 2006) and this is integral to decision support tool success (DeLone and McLean 2003).

Many factors influence the acceptance, or *adoption*, of a decision support tool. For the purposes of this chapter, it will be helpful to assume the context in which decision makers already have sufficient trust in the knowledge base[1] and believe that the underlying theory and technology can actually improve the quality of their decision making. In this regard, decision support tool acceptance and utilization depend on two factors: the *usefulness* of the tool, and its *ease of use* (Shibl, Lawley, and Debuse 2013). The first influential factor, *usefulness*, can be defined as the degree to which the tool is compatible with the real needs of decision makers and their belief that their issues and objectives are effectively addressed by the tool. In other words, how much can the design remedy real operational challenges faced by key stakeholders and remove friction from the decision-making process? The second factor concerns *ease of use*, that is, whether decision makers are comfortable in using the tool on a regular basis. Do decision makers believe that the decision support tool so captures and addresses their needs, that they are motivated to engage with and derive benefit from the tool?

### III. The Design of An Effective Decision Support Tool

While designers of decision support systems may intend to build a tool with effective characteristics that encourage adoption and sustain use, such achievements are far from automatic. The translation of decision context specifications into a solid tool that embeds advanced analytics can be a daunting task. The gap between theory and practice ensures routine encounters with practical, theoretical and technical limitations in the transformation of real-world problems into a decision support context. In many cases, assumptions and simplifications of the original problem must be weighed and specific techniques used to bridge this gap, so as to ensure the final outcome is as close as possible to the initial specifications of stakeholder needs. At the same time, the tool should have a compelling design that motivates stakeholders to routinely engage with it in their decision-making processes. In short, the benefits of gained analytical insights and ease of use should (far) outweigh the various costs such as opportunity, setup, training, and switching.

It is therefore critical to understand the behavioral and technical challenges of designing, developing, and implementing successful and effective decision support tools. There are a variety of approaches for designing and developing decision support tools and experts differ in opinion on what methodology works best. Regardless of the chosen methodology, we believe complementary skills and expertise are inherent for *successful implementation* of decision support tools, and these skill sets are just as important as the theory, knowledge and methodology used in their creation.

We maintain that the key to designing successful decision support tools is having deep understanding of the needs of key stakeholders together with compulsion to address these needs through incorporation, to the extent possible, in the tool. Without this, tool design and development take place from the limited perspective of the designer and developer, rather than the collective perspective that is inclusive of all stakeholders (Power 2002). This should take place through a comprehensive process of understanding stakeholder needs, which allows for effective collaboration through possibly extensive engagement to share ideas and brainstorm better design options. It is true that tool designers are experts in their respective technologies. At the same time, stakeholders are the experts in their own fields and their views should inform and drive the technology. While there is a time and place for designers to advocate their viewpoints and challenge those of others, we maintain that the act of carefully listening to the needs of stakeholders creates a perspective that can counteract the otherwise common tendency to rely on one's own judgment and sense of right design. After all, it is the stakeholder needs that are ultimately being addressed, for a successful implementation to take place. Hence, the goal is to build a collective point of view in which we complement rather than dominate one another's expertise.

---

[1] For more information on knowledge-based decision support systems, we refer interested readers to Chung, Boutaba, and Hariri (2016).

While designers have expertise in various analytics and sophisticated technologies, this does not supersede proper framing of what is to be developed. To understand this, stakeholders must know that designers have their best interests in mind. As Teddy Roosevelt once said, *People don't care how much you know, until they know how much you care*. Designers would do well to recognize that a fear common to humanity is that of becoming irrelevant or being replaced through technology or other means. This concern should be recognized by the design team and resolved at the inception of the design process, and stakeholders regularly reassured to reinforce the position. Wherever possible, stakeholders should know that the intentions of the tool are to assist, rather than replace.

To this end, we advocate that this process begins by asking open-ended (rather than direct) questions. Designers should employ active listening techniques such as paraphrasing responses from stakeholders, answering any questions that arise, and documenting their discussions. Designers can follow up initial questions to better motivate, guide, and draw out stakeholders to explain their missions and needs. Stakeholders will begin to recognize their needs are understood and that designers are there to assist them. This process encourages a continuous and productive dialogue that helps set expectations and structure stakeholder perspectives into an operational form amenable to the designer. In this way, a strong foundation and working relationship will be formed that allows designers to uncover the true needs of stakeholders.

Achieving an impactful decision support tool is a two-way street. Figure 1 depicts the usefulness of an iterative process to evaluate developmental, version-to-version milestones that engage stakeholders and incorporate their feedback. In particular, we have found that direct implementation of key primary needs of stakeholders via a minimal viable prototype of the tool, often yields productive developments and follow-up discussions that can later be smoothed in additional versions. This process allows stakeholders to get a firsthand view of an early stage prototype that ideally reflects their needs as well as shows both the promise and some extent of what the tool can deliver. If done well, this increases the trust with the design team, and the continuing evolution of understanding only improves the final product, deepens the level of engagement and trust, and increases the likelihood of designing a final version that stakeholders will adopt. When contrasted with single cycle development, where designers invest a great deal toward a near-final version based on an initial understanding without regular feedback, the iterative process eventually can result in significant savings of time, money and energy.

This process of iteratively listening, designing, engaging, and refining has the benefit of deepening the constructive communication between designers and stakeholders. As design proceeds, the topics under discussion can become more nuanced, which also increases the likelihood of misunderstanding and disagreement. Even so, these challenges can be alleviated by a combination of goodwill built up over previous iterations and prototypes, together with the practice of asking probing questions and active listening (Henrion 2019). This repeated process of asking, listening, clarifying, and understanding should continue throughout the course of the development.

Two critical aspects for the existence of any decision support tool are that stakeholders must have a complex problem that can be addressed with analytics, and that designers have a sufficient understanding of the required problem-solving analytics to be embedded within such a tool. Starting from this framework, we next illustrate through a series of vignettes three distinctive themes that can inform the design and influence stakeholders toward adoption of analytics-embedded decision support tools.

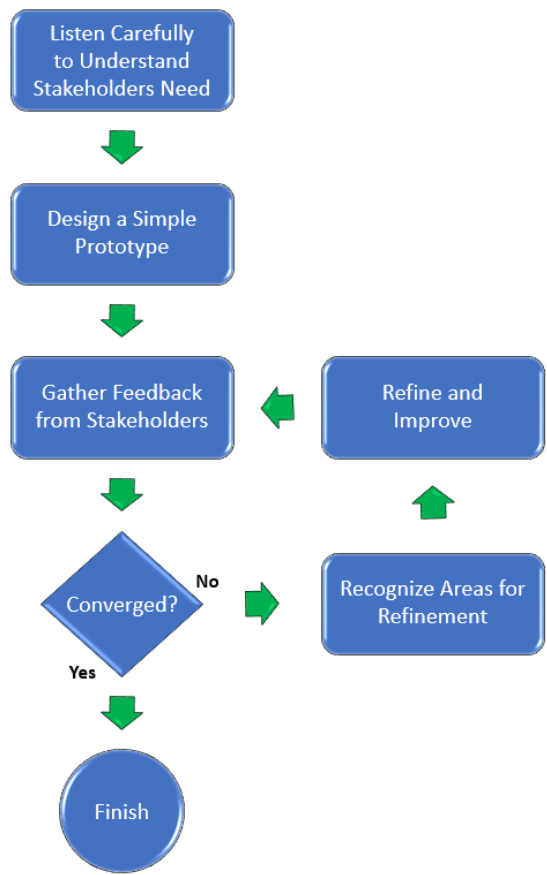

Figure 1: Iterative design process for human-centric decision support tools with active stakeholder engagement required for successful adoption.

From developing and maintaining strong relationships with stakeholders, to direct engagement of leadership for increased tool investment and buy-in, to the careful interactive interface design to elevate key decision-making interactivity with stakeholders, we maintain that these practices increase the likelihood of success. Done well, their incorporation can facilitate the best of both worlds: deep analytical insights, coupled with the ability for expert judgment to visualize, process, and perhaps interact with insights to arrive at final decisions, thereby increasing the likelihood of eventual tool adoption. While these themes appear in multiple vignettes, we feature a single most representative theme in each vignette.

**Vignette 1: The Global Opportunities Allocation Tool (GOAT)**

Recently, a need arose through engagement with the Global Experience Office (GEO) at Worcester Polytechnic Institute (WPI), who posed the following question: how can the annual process of manually allocating more than one thousand sophomore students to international project centers be redesigned? Can analytics be leveraged to account for preferences of students, alignment with individual project center needs, as well as the real-world constraints of the matchmakers?

The manual process that had been in place for many years required students to rank order project centers according to preference and promised the right to interview with the director of their number one rated center for the opportunity to secure a slot. For popular project center directors, this posed a significant challenge, interviewing at times greater than one hundred students that were eagerly competing for at most two dozen project center slots! Interviewing and then sorting through attributes among many qualified candidates was a major time sink and burden to the faculty. What's more, students that did not manage to be admitted into their top ranked project center ended up inevitably falling to a much lower slot for their eventual placement, as often their second, third and subsequent ranked locations also ended up being filled to capacity before the dust settled at a lower ranked project center. This left multiple secondary rounds of matching, and an overall less than favorable impression.

> *Key Insight:*
>
> *Develop & maintain strong working relationships with all problem stakeholders for sustained success.*

The process of inquiry in this context meant a series of meetings and surveys seeking answers to a variety of open-ended questions. With GEO serving as matchmakers, we were already regularly and iteratively engaging with their stakeholders. Throughout the development process, we additionally held focus groups for students to provide feedback on the survey instruments that elicited their preferences, promising gift card rewards in a random draw. For project center directors, we held focus groups to really listen and seek to understand their needs, rather than design according to what we assumed, or what was most convenient for us. By carefully listening to the actual needs of key stakeholders, including sophomore students, project center directors, and GEO matchmakers, we became familiar with the intricacies of all stakeholders and the overall matchmaking process.

Thereafter, one of the authors (Trapp) together with a team of WPI undergraduate students (Camila Dias, Lin Jiang, and Elizabeth Karpinski) built an analytical platform tool – known as the Global Opportunities Allocation Tool (GOAT, loosely named after WPI's caprine mascot, *Gompei*) – to conduct a large-scale matching of students to project centers according to student preferences and project center desires (Dias, Karpinski, and Jiang 2017). The GOAT takes the needs of project centers (such as languages spoken, writing skills, and leadership abilities) and both student preferences (including those centers to which students would not go) and attributes, formed match scores according to the alignment of students and eligible project centers, and matched in a manner that maximizes the overall score of matching every student to (eligible) capacitated project centers.

The GOAT generates visualizations to assist stakeholders in assessing and interacting with its recommendations. For each project center, the GOAT generates a view like that of Figure 2 detailing the levels of eight key attributes; Figure 2 depicts this for Reykjavik, Iceland. The bottom line (red) depicts the (relative) levels of the attributes desired by the project center, while the average (absolute) student scores for Reykjavik are in orange, and the average (absolute) student scores over all placements are in blue. Figure 3 depicts the number of students allocated, or *subscribed*, to various project centers. At a glance, decision-making stakeholders can review the recommended placements at a macro level, to understand whether project centers are full. Seeing such information at a glance is incredibly useful in making final decisions, for example such as deciding whether to open a significantly undersubscribed project center.

The end result was improved outcomes across the board. In the first year of use, a *full 100% of students* were placed in a center in which they were most interested, project center directors saw their need to interview vanish and their workload lessened, and GEO reduced the amount of manual paperwork and processing time overall by *approximately two months*. With all students placed to one of their top-ranked centers, and project center directors on the whole satisfied with the GOAT allocations, we continue to be engaged in the development of the GOAT for future years. Since 2017 the GOAT has been in regular use and has been featured favorably in the media (Fidrocki 2019). Moreover, ongoing work continues to further enhance its feature set, including research to ensure notions of fairness and diversity in the match outcomes, while maintaining efficient solution times (Wiratchotisatian, Atef Yekta, and Trapp 2021). We are grateful for the generous and collaborative support of GEO stakeholders that continue to appreciate the value and quality match outcomes generated by the GOAT over the former manual process bottlenecks.

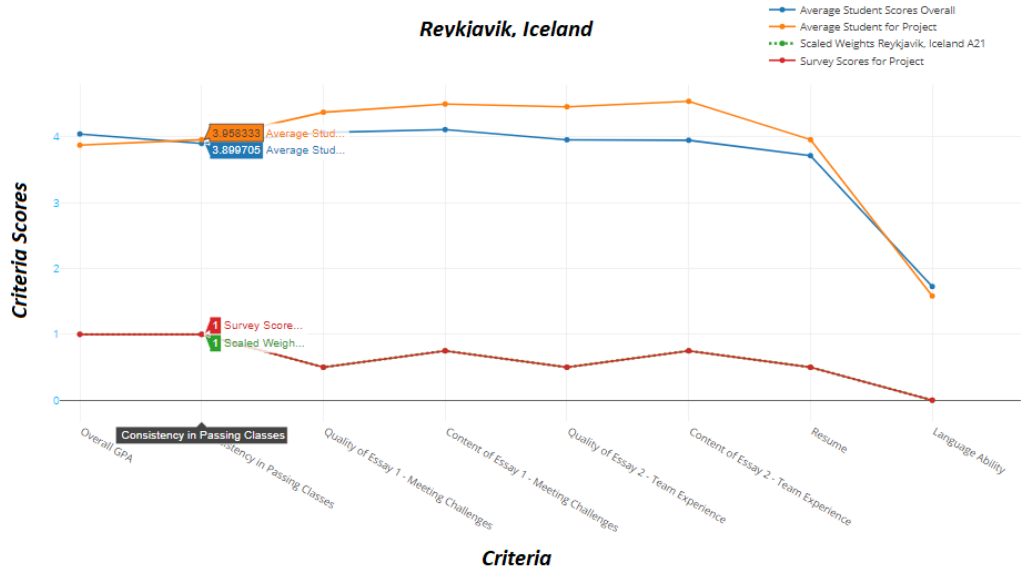

*Figure 2: Visualizing key attributes for matching students at the Reykjavik, Iceland project center.*

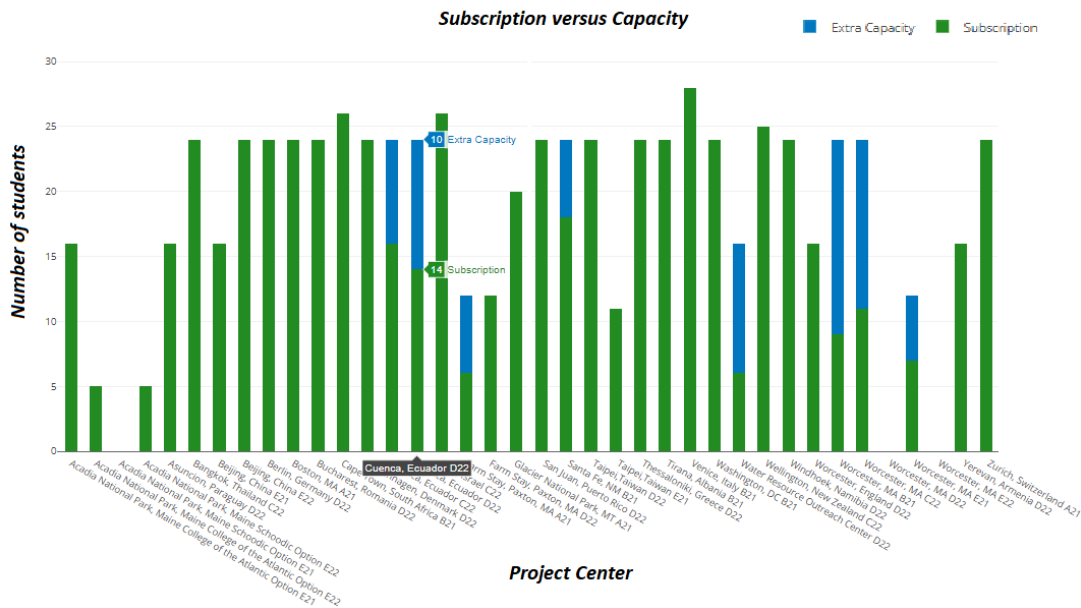

*Figure 3: Visualizing recommended student subscription, and remaining capacity, across various project centers.*

**Vignette 2: Optimizing the Scheduling of Asesores at Fundación Paraguaya (FP)**

Fundación Paraguaya (FP) is a multinational nonprofit and social enterprise headquartered in Asunción, Paraguay that develops innovative solutions to poverty and unemployment. Their initiatives include microfinance programs that loan funds to microentrepreneurs in small committees to elevate the socioeconomic status of participants. Its founder, Dr. Martin Burt, is Social Entrepreneur in Residence at Worcester Polytechnic Institute (WPI) and has partnered with WPI in a variety of projects and contexts. In early 2016 one of the authors (Trapp), together with students, engaged FP on embedding analytics into a decision support tool. An initial idea was to use optimization to design where wells could be placed to improve water access for impoverished populations. From an analytics perspective, the prospects seemed clear – sufficient data already existed to understand where water access was lacking. However, in discussions with Martin, a key insight emerged: if we only use analytics to inform where wells ought to be placed, then unless carefully done, their placement would be in vain. Even if in an optimal location, the poor would not own the well once placed; thus, when it inevitably degrades, it would likely be abandoned. Rather, designers need to understand how to engage the poor in the construction of wells; then they would own it. There is a parallel in decision-support tools: decision-making stakeholders need to be integrally vested in the work, so that the created tools are embraced and owned (Project Management Institute Inc. 2017).

> **Key Insight:**
> 
> Direct leadership buy-in and involvement increases the likelihood of success and eventual adoption.

Subsequent to this, we learned of a key need of FP social workers, or *asesores*. Asesores do the demanding work of FP's microfinance program: they find, connect, community build, assess, and hold accountable local decision-maker committees of impoverished individuals, typically comprised of women. We learned of an opportunity to improve the manner in which asesores manage the scheduling of their decision-maker committee appointments, which typically took place in what appeared to be ad hoc fashion throughout the week. Our initial thoughts centered on creating a decision support tool that embeds analytics for improved routing decisions for asesores in urban Asunción. Many good things took place from a visit to Asunción in January 2016. This enabled us to observe the situation on the ground and ascertain where analytics solutions could potentially benefit FP stakeholders. By traveling with asesores, we shadowed committee meetings and learned of their duties and needs. We visited their offices in the community and learned firsthand what pain points were. Only through careful interviews and leading questions, did we begin to understand the needs of key stakeholders, that is, the asesores and their supervisors. What was needed was something other than a routing application!

In this manner, we learned that a true need of asesores was a solution to their difficulty in keeping weekly scheduled appointments, which included navigating through busy Asunción. We thus realized asesores really needed a scheduling tool to assist with making schedules more efficient, instead of a routing tool. These insights led to the adoption of an analytics-based approach to group geographically adjacent committees via optimal clustering. We are confident that traveling to Asunción in-person enabled the necessary shadowing and interviewing to understand the situation on the ground and ascertain their true needs. Only then could we build an appropriate solution that addressed this problem. The initial solution was developed and delivered using VBA within Excel. Figure 4 shows the straightforward user interface that stakeholders could use to configure details such as how many daily meetings should be scheduled for asesores. Figure 5 details representative output of appointments that were clustered according to location, thereby minimizing travel throughout Asunción. Later, due to changing needs at FP, we were able to re-engineer the (backend) solution to be in the Java programming language, so as to make the schedule-recommender compatible with their existing database deployment.

| | | |
|---|---|---|
| Minimum Meetings per Day | 3 | Edit |
| Maximum Meetings per Day | 9 | Edit |
| Maximum Meeting Time per Day (min) | 360 | Edit |

| | |
|---|---|
| Number of Days | 5 |
| Number of Clients in Database | 22 |
| Number of Clients Selected | 15 |
| Number of Meetings Selected with Specified Time | 15 |
| Quantity of Duplicate Clients | 1 |

*Figure 4: User interface of asesore scheduling tool for Fundación Paraguaya.*

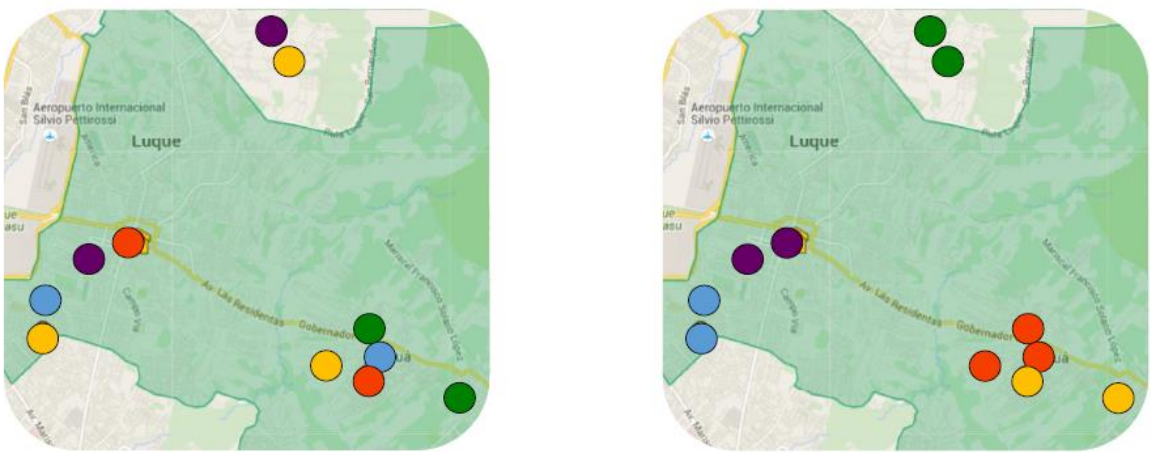

*Figure 5: Visualization of community meetings for asesores in Asunción, Paraguay. Left panel depicts existing meeting schedules, while the right panel shows the grouping that minimizes the need for traveling.*

**Vignette 3: Annie^TM MOORE**

Annie^TM MOORE (Matching and Outcome Optimization for Refugee Empowerment), named after the first immigrant on record at Ellis Island, New York in 1892, is an innovative and interactive decision support tool for refugee resettlement decision making (Ahani et al. 2020). Annie^TM assists the US resettlement agency HIAS (founded as the Hebrew Immigrant Aid Society) to find the best match between refugee families, or *cases*, and their network of communities, or *affiliates*. Annie^TM is a web application that combines techniques from integer optimization and machine learning to generate data-informed match recommendations, while at the same time giving substantial autonomy to resettlement staff to interactively fine tune the final recommendations.

We have developed Annie^TM in close collaboration with HIAS representatives from all levels, including leaders and practitioners. The key to the success of Annie^TM is our close working relationship with project stakeholders, which has been sustained and cultivated over the course of several years. Our level of rapport allows us to comfortably discuss and carefully analyze their day-to-day refugee resettlement process, so as to identify the operational challenges faced by resettlement staff and provide effective solutions for remedy. We intentionally designed Annie^TM in deference to HIAS decision makers - typically resettlement staff. Our match recommendations are suggestive in nature, and Annie^TM is built to empower decision makers to use their discernment and discretion in making final match decisions.

*Key Insight: Careful design of tool interface can reduce friction and improve interaction with stakeholders.*

Prior to Annie^TM, matching refugees to affiliates was largely done manually by pre-arrival resettlement staff at HIAS. This process had multiple sources of inefficiencies that motivated the development of Annie^TM. First, it is challenging to keep in mind all of the factors involved in the resettlement process. Refugee families have various need attributes that should be supported by their assigned affiliates. These attributes include language, nationality, and family composition. Manually matching refugee families into the communities in which their needs are met and at the same time respecting the allocated capacity of the affiliates is a challenging endeavor. Second, there are various established indicators that assess outcomes of refugee families in the affiliates. However, estimating welfare outcomes across affiliates is by no means trivial. Even assuming knowledge of welfare outcomes, humans are hard-pressed to manually find the best placement of refugee families into affiliates that respects available capacities.

Ample evidence in the literature suggests that refugee outcomes are greatly affected by the initial placement location (Åslund and Rooth 2007, Åslund and Fredriksson 2009, Åslund, Östh, and Zenou 2010, Åslund et al. 2011, Damm 2014). Societal integration is paramount for refugee outcomes, and a variety of indicators have been proposed to evaluate successful integration, including employment, housing, education, and health (Ager and Strang 2008). In the United States, the only widespread integration indicator available is the (binary) refugee employment status, as measured 90 days after arrival. We thus used advanced machine learning methods to derive signals from past refugee data and feed into an optimization model to generate data-driven, optimized recommendations on placements of refugees into communities. Our machine learning algorithms estimate the likelihood of refugee employment in affiliates. Using placement and outcome data on refugees arriving in past placement periods, we learn predictive models and then predict the employment probability of arriving refugees in subsequent placement periods. Even so, an algorithmically-driven decision support tool in the resettlement context is useful only to the degree to which it can be carefully evaluated by decision-making stakeholders. That is, with such high-stakes decisions involving vulnerable people, decision makers ought to be empowered in the matching process to have control over tool-generated recommendations. Therefore, *Annie*™ was designed as an interactive decision support tool so that resettlement staff can interact with all aspects of the problem context. We next describe the interactive design of *Annie*™.

Figure 6 depicts the *View Results* view of *Annie*™ which visualizes the optimal matching results. For any group of refugees, *Annie*™ identifies an optimal placement for refugee families (green tiles) throughout the network of HIAS affiliates (blue tiles) that maximizes the total family-affiliate match score, that is the total expected number of employed refugees. Family tiles are located in affiliate tiles to which they are optimally assigned, and their match scores are displayed in front of their names. At the top of the page the total expected number of employed refugees is displayed. Upon clicking on a refugee family tile, their unique attributes are displayed including language, nationality, and family composition (large family and single parent households). Similarly, clicking on affiliate tiles shows the available support services (Please see "Affiliate A" and "Case_22" at the top left of Figure 6). Affiliate tiles feature small arrows that allow for rapid adjustment of capacity for number of cases (C) and number of refugees (R).

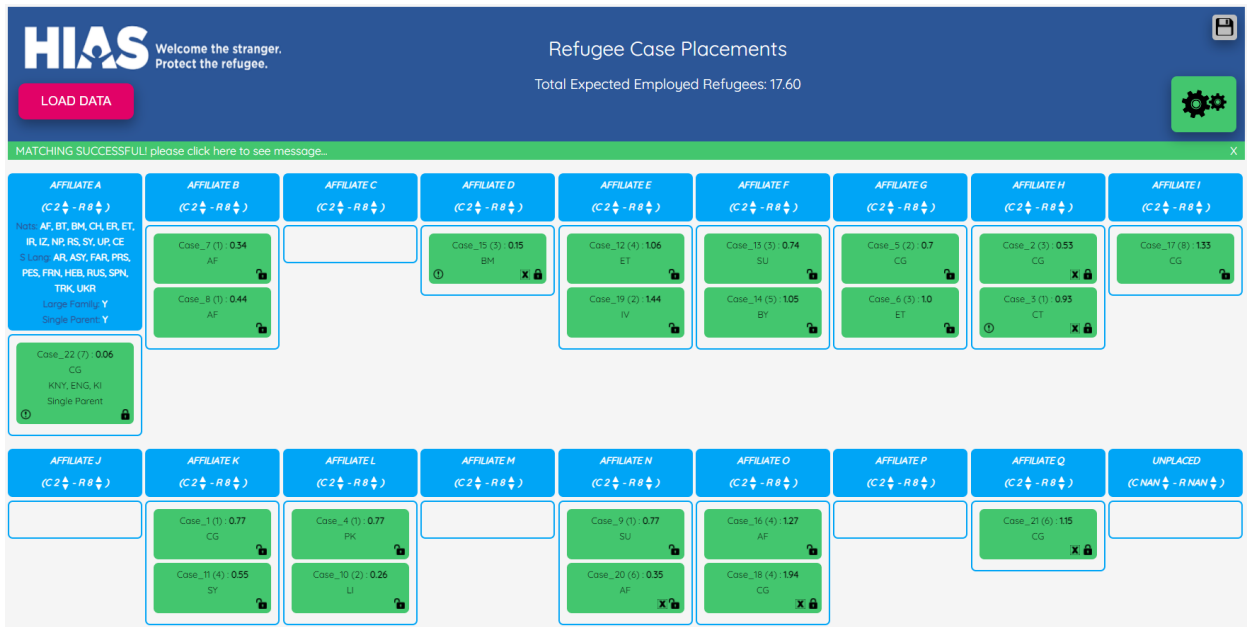

*Figure 6: The View Results view for the refugee resettlement software Annie™ MOORE. Here, decision-making stakeholders can easily interact with recommended family placements into communities in the HIAS network of affiliates.*

Interactivity is the most important characteristic of *Annie*™. Resettlement staff can interact with the optimization space and matching environment to change the placement recommendations and other settings until a desired assignment of refugee families to affiliates is found. Thus, decision makers may drag refugee family tiles to other affiliates, as shown in Figure 7, and when hovering over other affiliates the match scores dynamically update. Moreover, the total expected number of employed refugees also dynamically updates. In this way, decision makers can readily see the effect of moving refugee families to alternative affiliates at a glance.

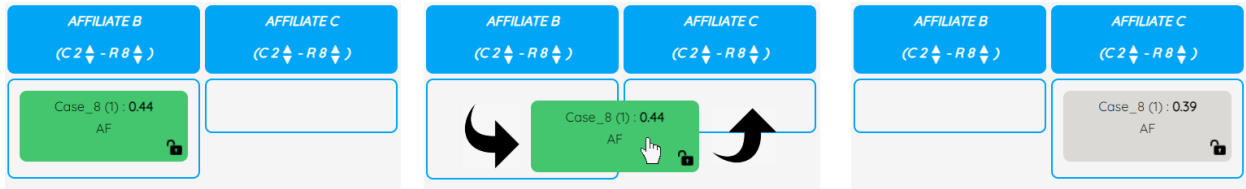

*Figure 8: Annie™ allows for seamless interaction with recommended placement via dragging family tiles to other communities.*

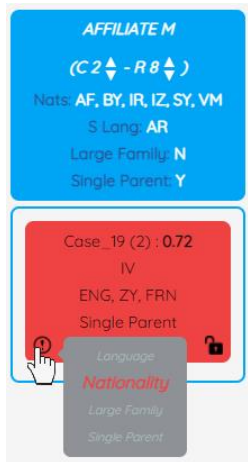

*Figure 7: Annie™ shows incompatible matches in red.*

*Annie*™ was carefully designed to incorporate useful color-coding to aid in decision making. Immediately after optimization, all refugee family tiles have a green background indicating an optimized state. Upon moving a family tile to another affiliate, the background color changes from green, to either gray indicating non-optimized (see Figure 7), or red indicating a mismatch in family-affiliate support services (see Figure 8). To gain further information on the mismatch(es), a small exclamation mark will appear on red family tiles; hovering over this icon will generate a list of any unsupported needs for that particular refugee family in that affiliate. Thus, at a glance decision maker can understand how families align in various affiliates.

Moreover, some refugee families are linked to other families or friends, or possibly affiliates; these are known as *cross-references*. It is desirable to place such families with their cross-references, so as to facilitate integration. Families with cross-references are visualized with a cross "X" icon on their tiles. Hovering over this icon reveals with a yellow border the cases and/or affiliates to which the family is cross-referenced. Figure 9 depicts an example where two refugee families are cross-referenced not only to one another (for example, adult siblings), but also to an affiliate.

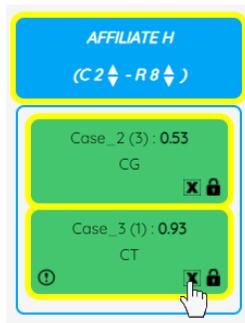

*Figure 9: Annie™ shows family member links in yellow.*

Still other features of *Annie*™ exist, such as the ability to *lock* an assignment into place. After moving family tiles to a desired affiliate, decision makers can *lock in* this match through clicking on the family tile lock icon. This can be seen in Figure 10 and is possible even for families and affiliates that are mismatched. Reoptimizing by clicking on the gray gear icon will override the mismatch as the design of *Annie*™ is to defer to the desires of the decision makers. For example, while there may not be an exact language match at an affiliate for a family, decision makers could know that another language that is spoken at that affiliate, is sufficiently similar and appropriate for placement.

More information on the design and implementation of *Annie*™ can be found in (Ahani et al. 2020). Through carefully designing the interface of *Annie*™ to be responsive to the needs of the decision maker, we have created a tool that combines the best of both worlds - advanced analytics, with powerful and easy to use interactive visualization that enables decision makers to fine-tune our recommendations to finalize match decisions.

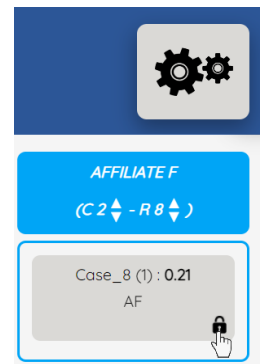

*Figure 10: Annie™ allows locking in adjusted placements, to re-match.*

## IV. Concluding Remarks

Decision-support tools are vehicles that use advanced analytics to bring deep insights to challenging decision problems into the hands of relevant stakeholders. Done well, decision-support tools have the potential to deliver significant value. We maintain that elevating the involvement and perspectives of key stakeholders beyond only the perspectives of those developing the analytics, will increase the overall value obtained and the likelihood of tool adoption.

Throughout this chapter we have explored factors that aid in increasing the value of analytics through well-designed decision-support tools. Advanced analytics, while a necessary ingredient, is increasingly insufficient to alone drive successful adoption of such tools. Through a series of vignettes detailing the collective experiences of the authors in designing effective decision-support tools, we maintain that factors which can increase the likelihood of tool acceptance and eventual adoption include the development and maintenance of strong relationships with problem stakeholders; direct stakeholder engagement and buy-in of the tool; a deep understanding, refined over time, of the actual stakeholder problem and its nuances; and the careful integration of key derived stakeholder needs and desires into a sleek interface that enables interaction with analytical insights for their processing and eventual consumption. Coupling solid analytics with decision support technologies that are built upon truly recognized needs of project stakeholders, is a powerful combination for decision support tools to be adopted and prosper.

**Acknowledgments**

We are grateful to the National Science Foundation (Operations Engineering) grant CMMI-1825348 for their generous support.